\documentclass[%
 pre,amsmath,amssymb,
]{revtex4-1}

\usepackage{graphicx}
\usepackage{dcolumn}
\usepackage{bm}

\usepackage[utf8]{inputenc}
\usepackage[T1]{fontenc}
\usepackage{mathptmx}

\newcommand\leftidx[3]{%
  {\vphantom{#2}}#1#2#3%
}

\begin{document}


\title[Time scales in systems with multiple exits]{Different time scales in dynamic systems with multiple exits}
\author{G. Bel}
 \altaffiliation[Also at ]{Center for Theoretical Biological Physics, Rice University,  Houston, TX 77005-1892, USA}
 \email{bel@bgu.ac.il}
 \affiliation{ 
Department of Solar Energy and Environmental Physics, BIDR, and Department of Physics, Ben-Gurion University of the Negev, Sede Boqer Campus 8499000, Israel\\
}
\author{A. Zilman}%
 \email{zilmana@physics.utoronto.ca}
\affiliation{ Department of Physics and Institute for Biomaterials and Bioengineering, University of Toronto, 60 Saint George St, M5S 1A7, Toronto, ON, Canada
}%
\author{A. B. Kolomeisky}%
 \email{tolya@rice.edu}
\affiliation{Department of Chemistry and Center for Theoretical Biological Physics, Department of Chemical and Biomolecular Engineering, Department of Physics and Astronomy, Rice University,  Houston, TX 77005-1892, USA}


\date{\today}

\begin{abstract}
Stochastic biochemical and transport processes have various final outcomes, and they can be viewed as dynamic systems with multiple exits. Many current theoretical studies, however, typically consider only a single time scale for each specific outcome, effectively corresponding to a single-exit process and assuming the independence of each exit process. But the presence of other exits influences the statistical properties and dynamics measured at any specific exit. Here, we present theoretical arguments to explicitly show the existence of different time scales, such as mean exit times and inverse exit fluxes, for dynamic processes with multiple exits. This implies that the statistics of any specific exit dynamics cannot be considered without taking into account the presence of other exits. Several illustrative examples are described in detail using analytical calculations, mean-field estimates, and kinetic Monte Carlo computer simulations. The underlying microscopic mechanisms for the existence of different time scales are discussed. The results are relevant for understanding the mechanisms of various biological, chemical, and industrial processes, including transport through channels and pores.
\end{abstract}

\maketitle

\section{\label{sec:intro}Introduction}

Many systems in chemistry, physics and biology operate in regimes in which a single input may result in multiple distinct outcomes. One example is nucleic acid synthesis, where chemically different sub-units can enter at the same positions for each newly created molecule.\cite{alberts2013essential,phillips2012physical,minchin2019understanding}
In this process, correct DNA and RNA molecules or molecules with mismatched nucleotides can be produced. 
Another example is the activation of T cells in the immune system.\cite{smith2009t,lever2014phenotypic,McKeithan1995,munsky2009,Bel2010,franccois2013phenotypic} A T cell that encounters a foreign peptide might undergo activation or remain quiescent depending on the molecular identity of the peptide. Conversely, in some cases, T cells might respond to a self-peptide, which can result in allergic reactions and autoimmune diseases.\cite{lever2014phenotypic} Another important example is the application of microfluidic devices for investigating chemical and biological systems.\cite{watanabe2017review} These devices utilize complex multi-channel structures for visualizing and controlling various processes. In these systems, multiple micro-channel exits are frequently utilized.  Furthermore, the translocation of molecules through channels and pores is crucial for many biological processes, and has been extensively studied, both theoretically and experimentally.\cite{Berezhkovskii2005,Berezhkovskii2003,Kustanovich2004,Zilman2007,Kolomeisky2007} All these processes can be viewed as dynamic systems with multiple exits.

Due to their considerable complexity, inferring the underlying molecular processes in these systems frequently relies on the indirect measurements of the exit dynamics at both the bulk and the single molecule levels.\cite{xing2019dynamic,lowe2010selectivity,english2006ever,nestorovich2002designed,danelon2006interaction} In  such systems, typically a single time scale is employed to describe both bulk and single molecule dynamics at the exit, ignoring the influence of other possible outcomes.\cite{english2006ever,reimann1999universal} However, the presence of other exits can affect the dynamics of the system, both spatially and temporally,\cite{Zilman2010,grinstead2012introduction} leading to the breakdown of the single time scale assumption. 
The goal of our investigation is to provide a rigorous theoretical framework for the quantitative study of complex dynamic processes with multiple possible outcomes.

It is shown in this paper that two independent time scales, a mean exit time and an inverse flux, are needed in order to fully characterize the exit dynamics. Both of them describe the statistics of exit events, but they behave differently when the kinetic parameters of the system are varied. The two time scales are the result of the presence of other exits in the system. To illustrate our theoretical arguments, we describe in detail three different dynamic systems, which are analyzed using exact analytical calculations, a mean field approximation, and kinetic Monte Carlo computer simulations. We show explicitly the existence of these time scales and their different dependencies on the system control parameters. The microscopic origin of the underlying processes is discussed.

\section{\label{sec:theor}Theoretical methods}

Consider a general dynamic process with $M$ possible outcomes. The process could be, for example, a system of a single enzyme molecule that may catalyze, in parallel, $M$ different substrates, producing $M$ different products $P_{i}$ ($i=1,2,...M$).\cite{alberts2013essential,phillips2012physical} In Fig. \ref{fig1}, we show a specific example of such systems with $M=2$ where the enzyme $E$ catalyzes two different processes, leading to the products $R$ (right product) and $W$ (wrong product). In our general explanations below, for convenience, we utilize the language of single enzymatic processes with multiple substrates, but our arguments are valid for all dynamic processes with multiple exits (or terminal states).

We start by assuming that the system has already reached the steady state, i.e., the total output flux is equal to the incoming flux, and $J_{i}$ is defined as a stationary current of the product $P_{i}$, where $i=1,2,...,M$. To characterize these processes, we also define $\Pi_{i}$ as a probability to reach the state $P_{i}$ for the first time before reaching any other product state starting from state $E$ (free enzyme). This exit probability is known as a splitting probability. \cite{van1992stochastic,redner-book} Similarly, we define a mean exit time $T_{i}$. This is a conditional mean first-passage time to reach the product $P_{i}$ starting from the free enzyme state.\cite{van1992stochastic,redner-book} Analyzing the dynamics of the system using a set of forward master equations allows us to evaluate explicitly the exit fluxes, $J_{i}$, in terms of the individual transition rates (see Fig. \ref{fig1}). The first-passage properties, $\Pi_{i}$ and $T_{i}$, can be evaluated using the backward master equations.\cite{van1992stochastic,redner-book} Our goal is to establish general relations between these dynamic properties of the system.

The total flux to make any product in the system is given by
\begin{equation}\label{eq1}
    J=\sum_{i=1}^{M} J_{i}.
\end{equation}
The mean time before the appearance of any of the products $P_{i}$ can be written as
\begin{equation}\label{eq2}
    T=\sum_{i=1}^{M} \Pi_{i} T_{i}.
\end{equation}
This expression emphasizes that this quantity is the average time over all possible outcomes, and the splitting probability $\Pi_{i}$ gives the probability that the system chooses the exit $i$. This total mean time and the total flux are related as
\begin{equation}\label{eq3}
    T=\frac{1}{J}, 
\end{equation}
which means that there is a single time scale for the overall production of {\it any} product in the system. However, such simple relations cannot be obtained for specific outcomes, the exit flux $J_{i}$, and the mean exit time $T_{i}$. Instead, one can write
\begin{equation}\label{eq4}
    J_{i}=\frac{\Pi_{i}}{T}.
\end{equation}
The physical meaning of this result is very clear: $1/T$ gives the frequency of making any of the product molecules, while $\Pi_{i}$ is the probability that this product is $P_{i}$. Together with Eq. \eqref{eq2}, this leads to 
\begin{equation}\label{eq5}
    \frac{1}{J_{i}}=\frac{\sum_{i=1}^{M} \Pi_{i} T_{i}} {\Pi_{i}}.
\end{equation}

Eq. \eqref{eq5} is our main result since it shows that there are two generally different time scales to characterize the exit dynamics, the inverse exit flux, and the mean exit time. These two times coincide only for a single-exit system ($M=1$). To quantify the deviations between different times scales, we define a parameter $R_{i}$,
\begin{equation}\label{eq6}
    R_{i}=\frac{T_{i}}{1/J_{i}},
\end{equation}
which is equal to one, only when both times are the same. Then from Eq. \eqref{eq5}, we obtain
\begin{equation}\label{eq7}
    \sum_{i=1}^{M} R_{i}=1.
\end{equation}
For example, for a simple system where all corresponding transition rates for all substrates are the same, it gives $R_{i}=1/M$. But generally, it can be shown that $0 < R_{i} <1$ (for $M>1$). 

Eqs. \eqref{eq6} and \eqref{eq7} imply that the mean exit time is always smaller than the inverse exit flux. The physical explanation of this observation is the following. The mean exit time, $T_i$, is the average time before the product $P_{i}$ is made  {\it after the last production event in the system}. But the last event is not necessarily a creation of the same product $P_{i}$ (i.e., it might be the creation of a different product, $P_{j\ne i}$). However, the inverse flux is exactly the average time between the appearances of the same product molecules. For this reason, we generally have $T_{i} < 1/J_{i}$. Thus, we predict that two different time scales must be employed to fully quantify the exit dynamics in complex systems with multiple outcomes. 

It is also important to note that the exit flux is a measure of the bulk properties of the system, i.e., it is the average over many cycles of the process and over many particles. But the mean exit time is the property of specific tagged particles. From this point of view, the output flux can be obtained from bulk dynamic measurements alone, while the mean exit times are determined from single-molecule measurements of labeled particles. Our theoretical analysis suggests that {\it both} types of experimental measurements are needed in order to fully characterize the dynamics and molecular mechanisms of systems with multiple exits.

\section{\label{sec:examples}Illustrative examples}

In order to better understand the microscopic origin of the existence of two different times scales for exit dynamics and its consequences for investigating real dynamic processes, we illustrate our theoretical arguments by considering three specific systems. In all of them, the dynamics can be analyzed by various means, thereby allowing us to clarify that the underlying physical principles may be reflected using different methods. 

\subsection{\label{subsec:kpr}Simple kinetic proofreading scheme}

\begin{figure}
    \includegraphics[width=\linewidth,trim=0cm 6cm 0cm 0cm,clip]{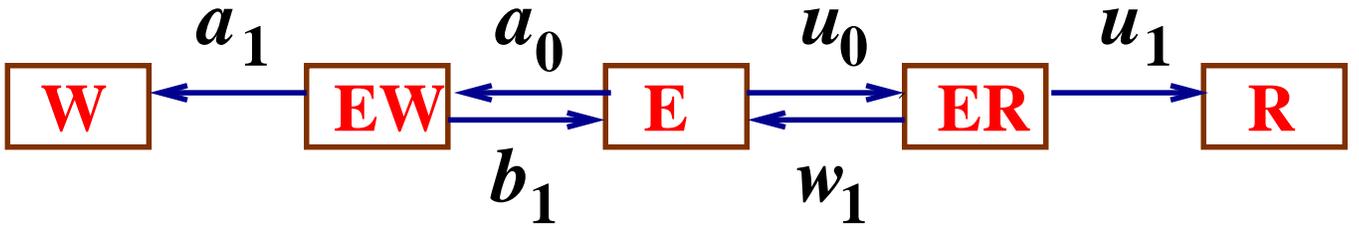}
    \caption{\label{fig1} A schematic view of the simple kinetic proofreading model, which can also be viewed as an enzymatic system with two substrates. Boxes describe different chemical states. The state $E$ corresponds to a free enzyme, the state $ER$ ($EW$) corresponds to the intermediate complex with the right (wrong) substrate, and $R$ ($W$) describes the right (wrong) product of the enzyme-catalyzed reactions. We consider that the system starts in the state $E$, and the possible outputs are the product states $R$ or $W$.}
\end{figure}

Let us start with a simple system shown in Fig. \ref{fig1}, where the enzyme molecule, $E$, can interact with two different substrates and produce two different products. This system can also be viewed as the simplest realization of kinetic proofreading mechanisms in biological systems, and the production of the right product $R$ competes with the production of the wrong product $W$.\cite{hopfield1974kinetic,johansson2008rate,banerjee2017accuracy} This is also the predominant view in explaining the mechanisms of T cell activation in the immune response.\cite{lever2014phenotypic,mckeithan1995kinetic}

From the free enzyme state $E$, the right substrate may associate with the enzyme with a rate $u_{0}$ to make the state $ER$, while the reverse reaction is characterized by a rate $w_{1}$: see Fig. \ref{fig1}. The right product $R$ is made with a rate $u_{1}$. Similarly, the wrong substrate can bind to the enzyme molecule with a rate $a_{0}$ to make the state $EW$, while the reverse reaction is characterized by a rate $b_{1}$: see Fig. \ref{fig1}. The wrong product $W$ is made with a rate $a_{1}$. Note also that although the rates $u_{0}$ and $a_{0}$ are viewed in our analysis as effectively unimolecular, in reality they are bimolecular and depend on the concentrations of right and wrong substrates, respectively.

We define the molecular fluxes to produce the right and wrong products as $J_{R}$ and $J_{W}$, respectively. The  probabilities for the system to make $R$ or $W$ are  described by the splitting probabilities $\Pi_{R}$ and $\Pi_{W}$, respectively. In addition, the mean exit times in the right and wrong directions are given by $T_{R}$ and $T_{W}$, respectively. These dynamic properties can be explicitly evaluated in terms of the individual transition rates, as explained in Appendix \ref{app:kpr}.
Obviously, the steady state output flux is non-vanishing only when there is an incoming flux of free substrates. However, the steady state ensures that the total output flux is equal to the incoming flux, which enables the elimination of the incoming flux from the expressions for the output fluxes.

The explicit expression for the inverse molecular flux for the $R$ molecules is given by,
\begin{equation}\label{eq15}
    \frac{1}{J_{R}}=\frac{(u_{1}+w_{1})(a_{1}+b_{1})+u_{0}(a_{1}+b_{1})+a_{0}(u_{1}+w_{1})}{u_{0}u_{1}(a_{1}+b_{1})},
\end{equation}
while for the $W$ molecules, we have,
\begin{equation}\label{eq16}
    \frac{1}{J_{W}}=\frac{(u_{1}+w_{1})(a_{1}+b_{1})+u_{0}(a_{1}+b_{1})+a_{0}(u_{1}+w_{1})}{a_{0}a_{1}(u_{1}+w_{1})}.
\end{equation}

Now, as shown in Appendix A, the splitting probability and the mean exit time in the $R$ direction are given by
\begin{equation}\label{eq24}
    \Pi_{R} =\frac{u_{0}u_{1}(a_{1}+b_{1})}{u_{0}u_{1}(a_{1}+b_{1})+a_{0}a_{1}(u_{1}+w_{1})};
\end{equation}
and
\begin{equation}\label{eq25}
    T_{R}  = 
     \frac{(a_{1}+b_{1})(u_{0}+u_{1}+w_{1})+a_{0}b_{1}\frac{u_{1}+w_{1}}{a_{1}+b_{1}} +a_{0}a_{1}}{u_{0}u_{1}(a_{1}+b_{1})+a_{0}a_{1}(u_{1}+w_{1})}.
\end{equation}
For the product $W$, we obtain,
\begin{equation}\label{eq26}
    \Pi_{W} =\frac{a_{0}a_{1}(u_{1}+w_{1})}{u_{0}u_{1}(a_{1}+b_{1})+a_{0}a_{1}(u_{1}+w_{1})};
\end{equation}
and
\begin{equation}\label{eq27}
    T_{W}=\frac{(u_{1}+w_{1})(a_{0}+a_{1}+b_{1})+u_{0}w_{1}\frac{a_{1}+b_{1}}{u_{1}+w_{1}} +u_{0}u_{1}}{u_{0}u_{1}(a_{1}+b_{1})+a_{0}a_{1}(u_{1}+w_{1})}. 
\end{equation}

\begin{figure}[htb]
    \includegraphics[width=\linewidth]{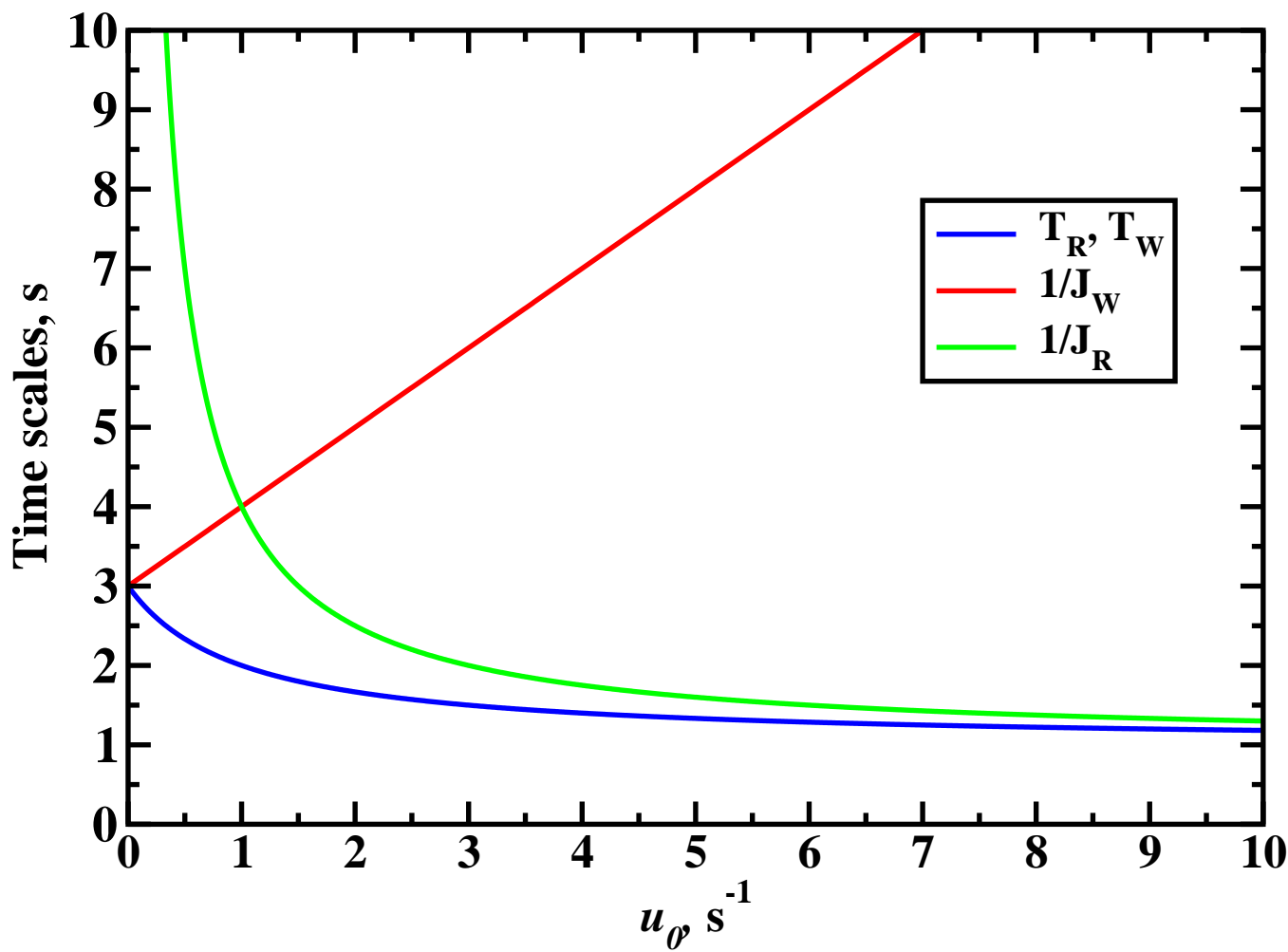}
    \caption{\label{fig2} Different time scales for the simple kinetic proofreading model as a function of the transition rate $u_{0}$. For the calculations, we used the following values for the transition rates: $u_{1}=w_{1}=a_{0}=a_{1}=b_{1}=1$ s$^{-1}$. Please note that for these values of the transition rates, the mean exit times, $T_{R}$ and $T_{W}$, coincide.}
\end{figure}

Comparing Eqs. \eqref{eq15} and \eqref{eq16} with Eqs. \eqref{eq25} and \eqref{eq27}, it can be shown that
\begin{equation}
    \frac{1}{J_{R}}=T_{R}+\frac{\Pi_{W}}{\Pi_{R}} T_{W};
\end{equation}
\begin{equation}
    \frac{1}{J_{R}}=T_{W}+\frac{\Pi_{R}}{\Pi_{W}} T_{R}.
\end{equation}
To emphasize that the time scales' behavior for each exit ($T_{R}$ and $1/J_{R}$, $T_{W}$ and $1/J_{W}$, respectively) may be very different, in Fig. \ref{fig2}, we present the dependence of these quantities on the transition rate $u_{0}$, while all other transition rates are the same. This corresponds to a situation where the concentration of the right substrates in the system is varied. 

One can see from Fig. \ref{fig2} that the mean exit for the right products and the inverse flux for $R$ generally are different quantities. They are essentially the same in the limit of $u_{0} \gg 1$ because, in this case, only the formation of $R$ molecules is possible, transforming the system into an effective single-exit process. However, the deviation between $T_{R}$ and $1/J_{R}$ starts to grow for decreasing values of $u_{0}$. In the limit $u_{0} \rightarrow 0$, the production of $R$ almost stops and $1/J_{R} \rightarrow  \infty$, while the mean exit time for those rare situation when the system goes in the direction of right products is still finite. 

The difference in the time scales' behavior for exiting in the wrong directions, $T_{W}$ and $1/J_{W}$, is even more striking. While the mean exit time $T_{W}$ decreases for larger transition rates $u_{0}$, the exit flux $J_{W}$ decreases and the corresponding time scale $1/J_{W}$ increases. In the limit of large $u_{0}$, only $R$ molecules are preferentially produced, and it takes many production cycles to produce occasionally the $W$ molecule. However, if the system goes in the wrong direction ($W$ is produced), it should happen relatively quickly (measuring the time since the free enzyme state, $E$). Both time scales in the $W$ direction are the same only in the limit of small $u_{0}$, when the system is biased toward the wrong direction. This means that again, the system effectively works like a single-exit process. 

Clearly, the differences between time scales for the same exit are due to the presence of the second exit, and they disappear in the regime where the system behaves as a single-exit process. The important conclusion from our theoretical calculations here is that a single time scale is not sufficient to determine the molecular mechanisms of a process with multiple exits. Two dynamic scales have to be utilized for each exit, and this again suggests that both bulk measurements and single-molecule studies must be employed in the analysis of complex dynamic processes.

\subsection{\label{subsec:2s}Exact solutions for a channel with two sites and two exits}

\begin{figure}
    \includegraphics[width=\linewidth,trim=4cm 16cm 3cm 2cm,clip]{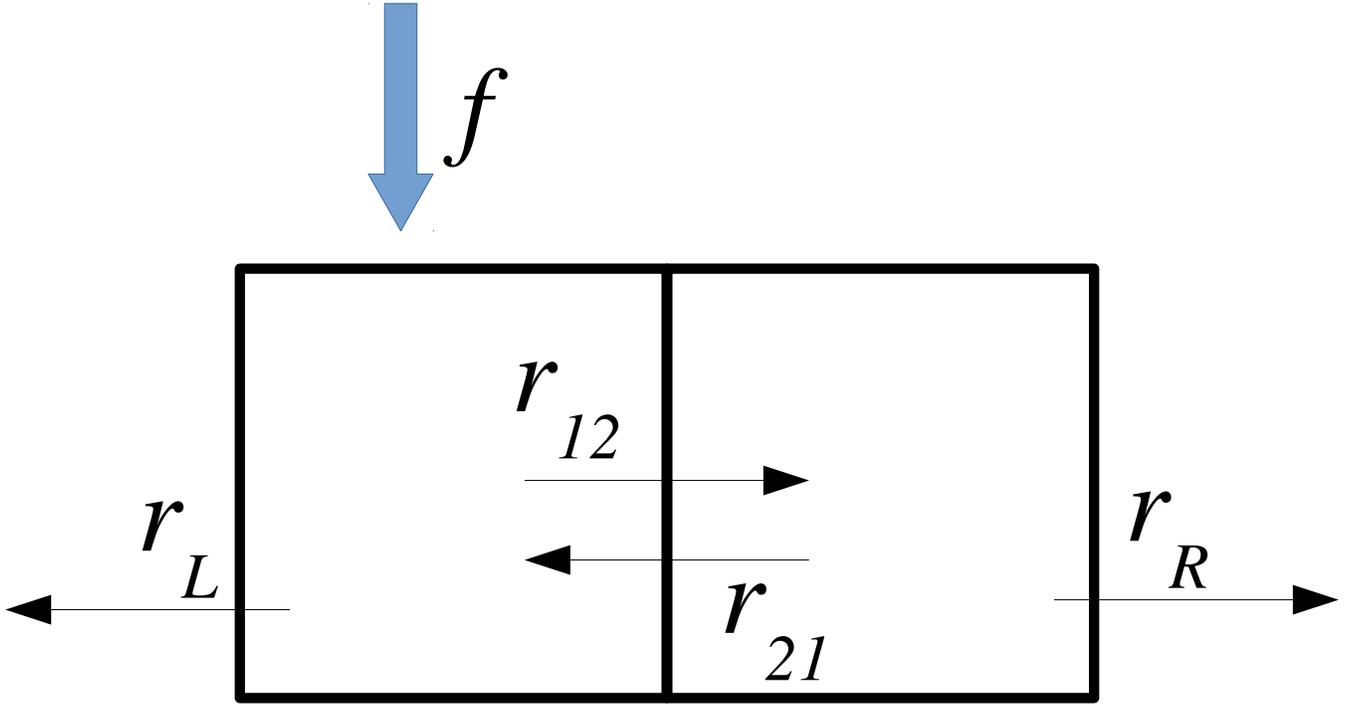}
    \caption{\label{fig:2s_schem} A schematic illustration of the two-site model. The arrows denote the possible transitions, and the label next to each arrow denotes the rate for the corresponding transition if the target is available (each site may be occupied by no more than one particle).}
\end{figure}

The next system to be considered here is a simple two-site channel model with exclusion, which can be viewed as the simplest realization of a complex dynamic system as found, e.g., in microfluidic devices. In our model, presented in Fig \ref{fig:2s_schem}, each site is either occupied by one particle or it is empty. The incoming flux inserts particles into the first site if it is empty, with a rate $f$; from there, the particle may move to the second site (if it is empty), with a rate $r_{12}$, or it may exit to the left with a rate $r_L$. From the second site, the particle may move back to the first site (if it is empty) with a rate $r_{21}$, or exit to the right with a rate $r_R$. A schematic description of the model is given in Fig. \ref{fig:2s_schem}. A similar system was considered in \cite{chou1999kinetics}, but only the flux was calculated. 

The system has four different states. We denote these states as $00$ when the two sites are empty, $10$ when the first site is occupied and the second is not, $01$ when the second site is occupied and the first one is not, and $11$ when both sites are occupied. In Appendix \ref{app:2s}, the full mathematical description of the dynamics in the system is provided. It is found that the steady state probabilities (${t\to\infty}$), in terms of individual transition rates for each of four states in the system, are:
\begin{align}\label{eq:2sss}
    \leftidx{^{ss}}{p}{_{00}}&=N_{2s}\left(\frac{(r_L+r_R) (r_{12} r_R+r_{21} r_L+r_L r_R)}{f^2 r_{12}}+\frac{r_L r_R}{f r_{12}}\right), \nonumber\\
    \leftidx{^{ss}}{p}{_{10}}&=N_{2s}\frac{f r_{R}+r_{21} r_{L}+r_{21} r_{R}+r_{L} r_{R}+r_{R}^2}{f r_{12}}, \nonumber\\
    \leftidx{^{ss}}{p}{_{01}}&=N_{2s}\frac{r_L+r_R}{f}, \nonumber\\
    \leftidx{^{ss}}{p}{_{11}}&=N_{2s}, 
\end{align}
where $N_{2s}$ is given by:
\begin{widetext}
\begin{equation}\label{eq:2sssN}
     N_{2s}=\frac{f^2 r_{12}}%
     {f^2\left(r_{12}+r_{R}\right)%
     +f \left(\left(r_{12}+r_{21}\right)\left(r_{L}+r_{R}\right)+r_{R}\left(2 r_{L}+r_{R}\right)\right)%
     +\left(r_{L}+r_{R}\right) \left(r_{R}\left(r_{12}+r_{L}\right)+r_{21} r_{L}\right)}.
\end{equation}
\end{widetext}
These expressions allow us to evaluate the  steady state fluxes to the right or to the left,
\begin{align}\label{eq:2sJ}
    J_R=r_R\left(\leftidx{^{ss}}{p}{_{01}}+\leftidx{^{ss}}{p}{_{11}}\right)\nonumber \\
    J_L=r_L\left(\leftidx{^{ss}}{p}{_{10}}+\leftidx{^{ss}}{p}{_{11}}\right).
\end{align}

The mean escape times to the right and to the left can be calculated using the backward master equations. Any particle entering site 1 either finds site 2 occupied or not. Therefore, the mean exit time is written as the appropriate average of these two initial conditions. The details of the calculations using the backward master equations are provided in Appendix \ref{app:2s}.
The result for the mean exit time to the right is given below:
\begin{equation}\label{eq:2sTR}
T_R=T_{R,2}\frac{\Pi_{R,2}p_{10}}{\Pi_{R,2}p_{10}+\Pi_{R,1}p_{11}}+T_{R,1}\frac{\Pi_{R,1}p_{11}}{\Pi_{R,2}p_{10}+\Pi_{R,1}p_{11}}.%
\end{equation}
In this expression, the factor $\Pi_{R,1}$ is the right exit probability, when the tagged particle is initially at site 1 and site 2 is empty. It can be written as
\begin{equation}\label{eq:2spir1}
\Pi_{R,1}=\frac{r_{12}r_{R}^{2}(f+r_{L}+r_{R})}{(r_{L}+r_{R})\left(
r_{21}r_{L}(r_{L}+r_{R})+(r_{L}+r_{12})r_{R}(f+r_{L}+r_{R})\right)  }
\end{equation}
The corresponding mean right exit time is:
\begin{widetext}
\begin{align}\label{eq:2sTR1}
T_{R,1}  & =\frac{\left[  \left(  r_{L}+r_{R}\right)\left(  r_{L}^{2}+3r_{L}%
r_{R}+r_{R}^{2}+r_{21}\left(  2r_{L}+r_{R}\right)  +r_{12}\left(  r_{L}%
+2r_{R}\right)  \right)  \right]  }{(f+r_{L}+r_{R})
\left(  r_{21}r_{L}\left(  r_{L}+r_{R}\right)  +\left(  r_{L}+r_{12}\right)
r_{R}\left(  f+r_{L}+r_{R}\right)  \right)  }%
+\frac{f  \left[  r_{21}\left(  3r_{L}%
+r_{R}\right)  +2r_{12}\left(  r_{L}+2r_{R}\right)  +2\left(  r_{L}^{2}%
+3r_{L}r_{R}+r_{R}^{2}\right)  \right]  }{(f+r_{L}+r_{R})  \left(  r_{21}r_{L}\left(  r_{L}+r_{R}\right)  +\left(
r_{L}+r_{12}\right)  r_{R}\left(  f+r_{L}+r_{R}\right)  \right)  }\nonumber\\
& +\frac{f^{2}\left[  r_{L}^{2}+3r_{L}r_{R}+r_{R}^{2}+r_{12}\left(
r_{L}+2r_{R}\right)  \right]  }{\left(  r_{L}+r_{R}\right)  \left(
r_{21}r_{L}\left(  r_{L}+r_{R}\right)  +\left(  r_{L}+r_{12}\right)
r_{R}\left(  f+r_{L}+r_{R}\right)  \right)  (f+r_{L}+r_{R})}.
\end{align}
\end{widetext}
Similarly, for the initial state when the tagged particle is at site 1 and site 2 is occupied, the right exit probability is:
\begin{equation}\label{eq:2spir2}
\Pi_{R,2}=\frac{r_{12}r_{R}(f+r_{L}+r_{R})}{r_{R}\left(  r_{12}+r_{L}\right)
(r_{L}+r_{R}+f)+r_{L}r_{21}(r_{L}+r_{R})}.%
\end{equation}
The corresponding mean right exit time is:%
\begin{widetext}
\begin{align}\label{eq:2sTR2}
T_{R,2}  & =\frac{f^{2}(r_{12}+r_{L}+r_{R})}{(f+r_{L}+r_{R})(r_{21}r_{L}%
(r_{L}+r_{R})+r_{R}(r_{12}+r_{L})(f+r_{L}+r_{R}))}
+\frac{f(2r_{12}(r_{L}+r_{R})+2(r_{L}+r_{R})^{2}+r_{21}(2r_{L}+r_{R}%
))}{(f+r_{L}+r_{R})(r_{21}r_{L}(r_{L}+r_{R})+r_{R}(r_{12}+r_{L})(f+r_{L}%
+r_{R}))}\nonumber\\
& +\frac{(r_{L}+r_{R})^{2}(r_{12}+r_{21}+r_{L}+r_{R})}{(f+r_{L}+r_{R}%
)(r_{21}r_{L}(r_{L}+r_{R})+r_{R}(r_{12}+r_{L})(f+r_{L}+r_{R}))}.%
\end{align}
\end{widetext}

\begin{figure}
    \includegraphics[width=\linewidth,trim=2cm 0cm 3cm 1cm,clip]{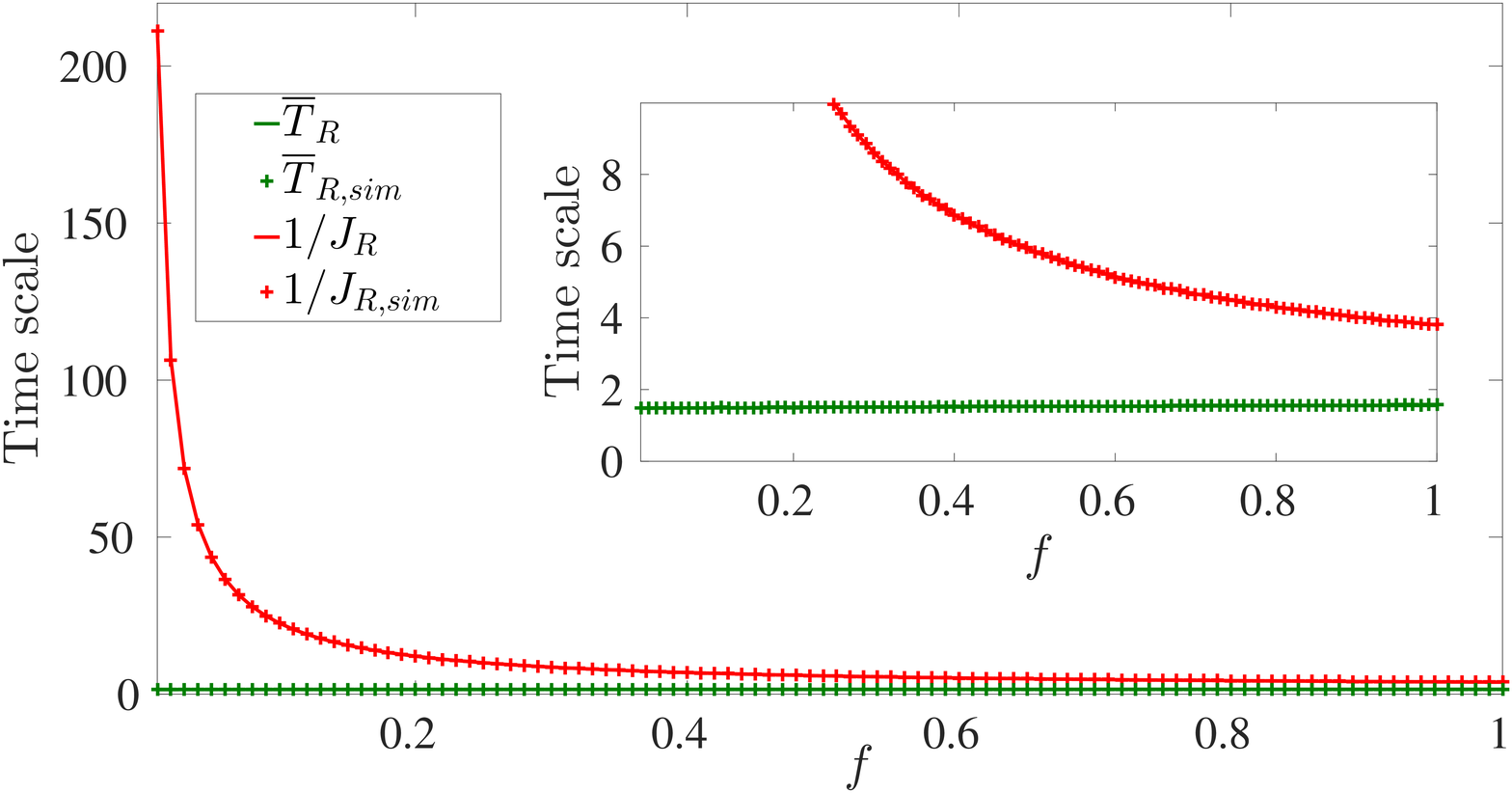}%
    \caption{\label{fig:2s_Time} The time scales in the two-site system. The lines depict the analytical results, and the symbols depict the corresponding simulation results. It is obvious that the time scale derived from the output current is very different from the time scale characterizing the mean escape time of tagged particles. The parameters used are: $r_{12}=r_L=r_R=1$ and $r_{21}=0.1$. The inset shows a reduced scale of the characteristic time axis.}
\end{figure}

In Fig. \ref{fig:2s_Time}, we present the mean right exit time and the inverse of the right output current against the incoming flux rate, $f$, for a specific set of parameters. The analytical results are compared with Monte Carlo computer simulations of the process. As expected, the simulations agree perfectly with the exact analytical solutions. The output current monotonically increases with an increase in the incoming flux rate until it saturates in the limit of the fully occupied system (i.e., when the first site is filled immediately after it becomes empty). Consequently, the inverse of the output current decreases monotonically. Eq. \eqref{eq:2sasymJ} in Appendix \ref{app:2s} describes the asymptotic limit of the output currents.

The mean first-passage time (or the completion time) is different than the time scale obtained from the steady state output current. For small values of the incoming current, $f$, the difference can reach orders of magnitude. In experiments, the mean exit time typically obtained from single particle measurements and the output current is determined from the bulk measurements. The difference between these two time scales emphasizes the need to combine the two types of measurements in order to properly characterize any dynamic system.

\subsection{\label{subsec:chan}General multi-site channel}

\begin{figure}
    \includegraphics[width=\linewidth,trim=1cm 16cm 1cm 6cm,clip]{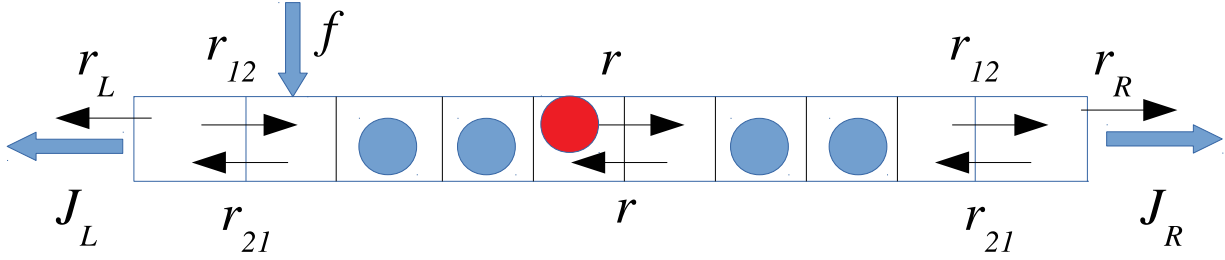}%
    \caption{\label{fig:ChanMod} Schematic description of the channel considered.
    The arrows represent the possible transitions (if the target site is not full), and the corresponding labels clarify the notation used for the various rates. See the text for more details.}
\end{figure}

In this example, let us consider a more complex system that describes the transport in a multi-site channel with two exits as presented in Fig. \ref{fig:ChanMod}. The input flux enters the second site with a maximal rate $f$ (when the site is not fully occupied), and the particles can hop within the channel in both directions. There are two exits, on the right and on the left ends of the channel. The transition rate from site $2$ to site $1$ (and, by symmetry, also from site $N-1$ to site $N$) is $r_{21}$, and the transition rate from site $1$ to site $2$ (and, by symmetry, also from site $N$ to site $N-1$) is $r_{12}$. The transition rates within the channel are assumed to be symmetric and equal to $r$ (in each direction): see Fig. \ref{fig:ChanMod}. In addition, the transition rates out of the channel are $r_R$ and $r_L$ to the right and to the left, respectively. Each site may be occupied by up to $m$ particles simultaneously ($m=1$ corresponds to exclusion process). The different rates at the ends of the channel are considered because the dynamics out of the channel and in the vicinity of the ends may be different from the dynamics within the channel. This model is analyzed using mean-field calculations supported by kinetic Monte Carlo computer simulations.

The analysis of the dynamic properties of the system consists of two stages. The first one describes the steady state population distribution in the channel. In the second stage, the dynamics of a tagged particle, assuming that the population distribution corresponds to the steady state distribution, is obtained. The equations describing the dynamics of the site population densities are provided in Appendix \ref{app:gc}. These equations can be solved using a mean-field approach in the steady state. 
Let us define $n_{k}^{ss}$ as the stationary occupation of the site $k$. The steady state solution for the internal sites, $2\leq k\leq N-1$ can be written as:
\begin{align}
n_k^{ss}/m&=1-A+kB, \ \ \ \ 2\leq k \leq N-1, \label{eq:sspan}
\end{align}
where the procedure to evaluate the variables $A$ and $B$ is explained below. Solving the steady state equations for the four boundary sites ($1$, $2$, $N-1$ and $N$) yields the steady state populations of the end sites, $n_1^{ss}$ and $n_N^{ss}$, in terms of the parameters $A$ and $B$,
\begin{align}\label{eq:sspan2}
&n_1^{ss}/m=\frac{r_{21}(1-A-2B)}{r_L+r_{12}(A-2B)};\\
&n_N^{ss}/m=\frac{r_{21}\left(1-A-(N-1)B\right)}{r_R+r_{12}\left(A-(N-1)B\right)}.\nonumber
\end{align}
The variables $A$ and $B$ are obtained by solving the following equations,
\begin{widetext}
\begin{align}\label{eq:sspan3}
r_{12}\frac{r_{21}(1-A-2B)}{r_L+r_{12}(A-2B)}&=\left(r_{21}+r\left(A-3B\right)\right)\frac{1-A+2B}{A-2B}-r\left(1-A+3B\right)-f/m; \\
r_{12}\frac{r_{21}\left(1-A-(N-1)B\right)}{r_R+r_{12}\left(A-(N-1)B\right)}&=\left(r_{21}+r\left(A-(N-3)B\right)\right)\frac{1-A+(N-2)B}{A-(N-1)B}-r\left(1-A+(N-2)B\right). \nonumber
\end{align}
\end{widetext}

It is important to note that due to the asymmetry of the transition rates at the end sites, the equations are not linear (i.e., the equation for site 1 involves the product of the populations of sites 1 and 2,  and so on). Therefore, our solution is only a mean-field approximation and not the exact solution. However, direct simulations of the process reveal that the mean-field and the exact steady state densities are very close for a large range of parameters.

The steady state population provides the exit currents in both directions as:
\begin{align}
 J_R=r_Rn_N^{ss};\nonumber \\
 j_L=r_Ln_1^{ss}.
\end{align}

\begin{figure*}[htb]
 \includegraphics[width=0.45\linewidth,trim=1cm 0cm 0cm 0cm,clip]{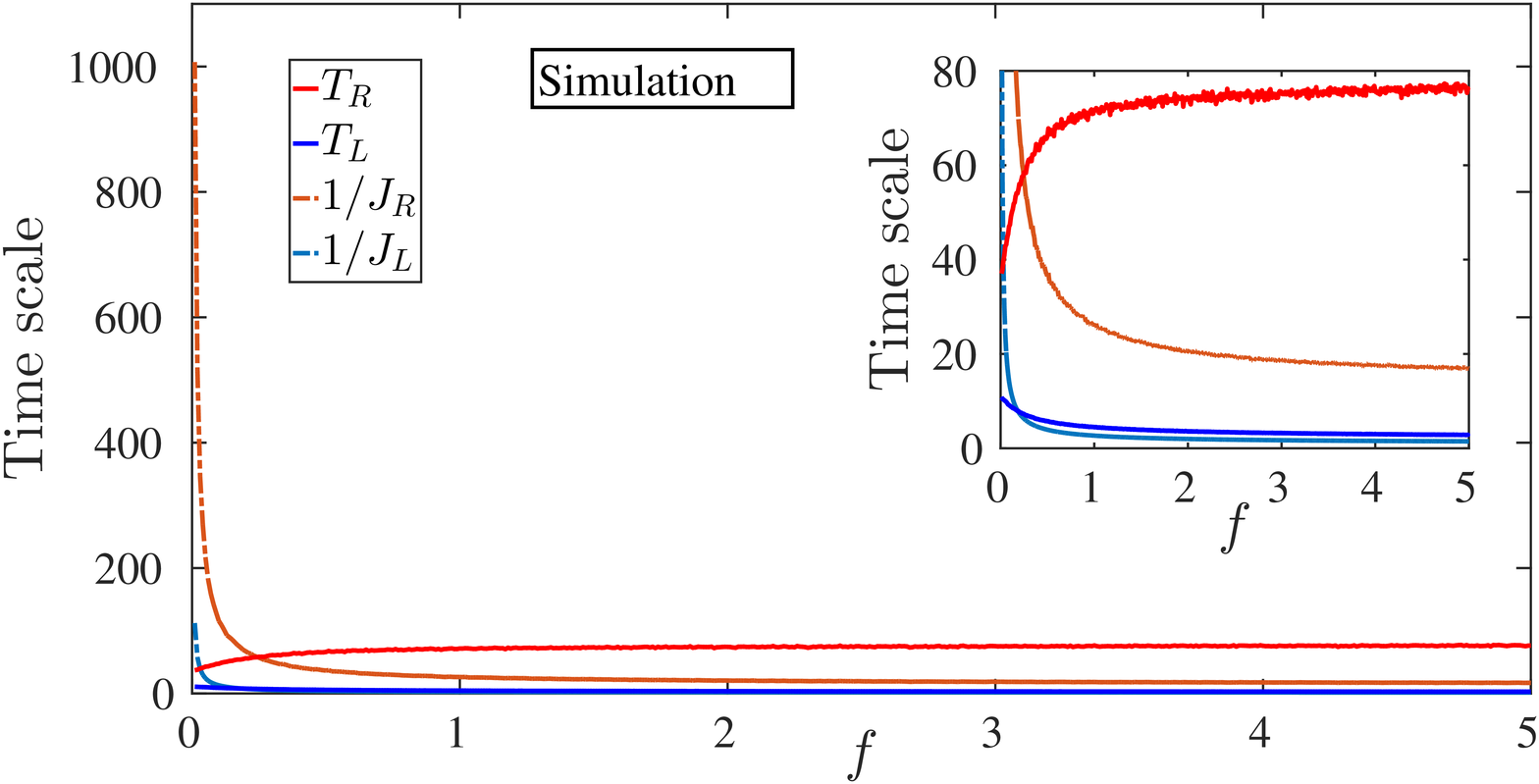}
 \includegraphics[width=0.45\linewidth,trim=1cm 0cm 0cm 0cm,clip]{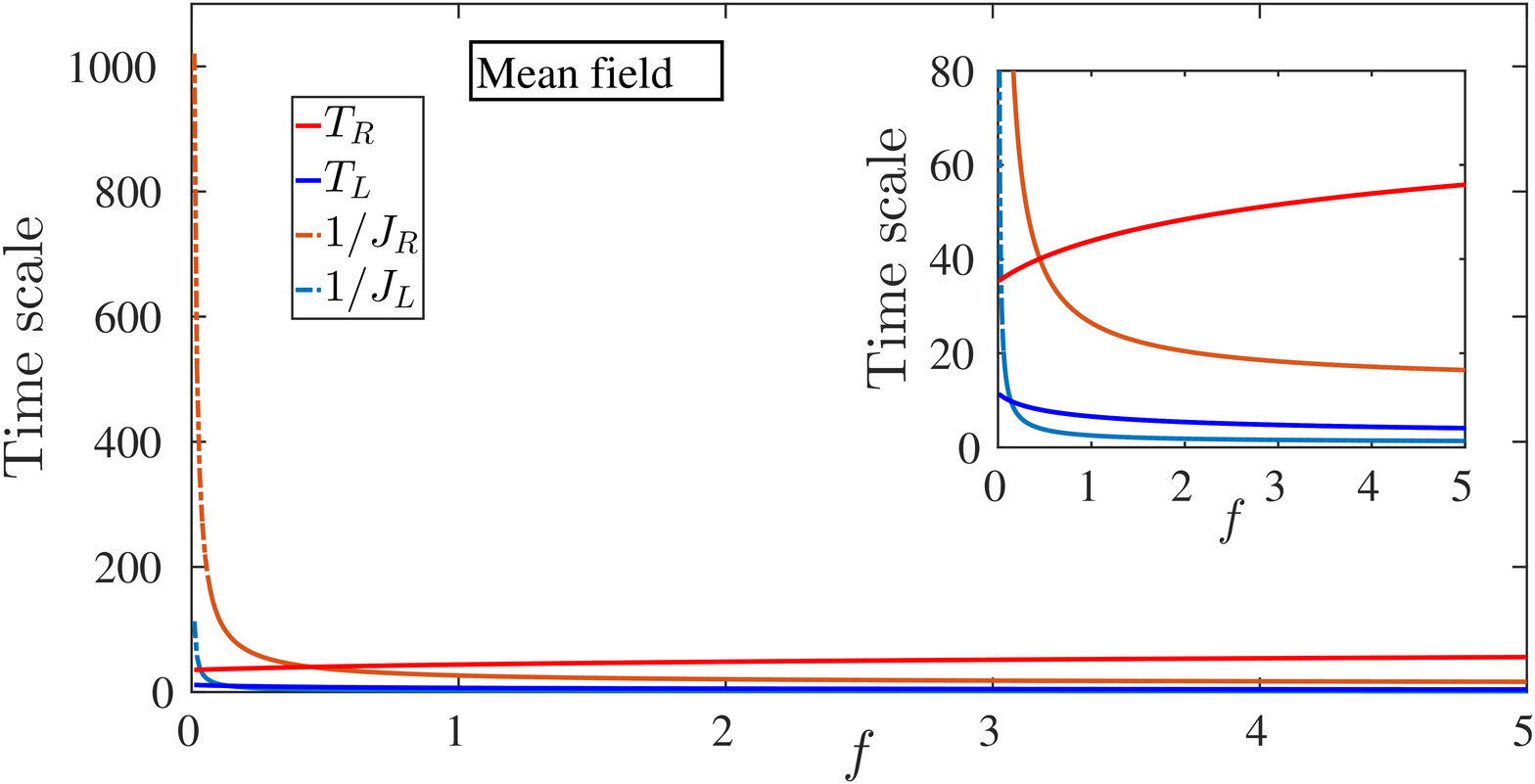}
 \caption{\label{fig:TimesGCrR0p1} The simulation (left) and mean field (right) time scales vs. the impinging current for the general channel (see Fig. \ref{fig:ChanMod}). The solid red and blue lines show the right and left mean escape times, respectively. The dashed-dotted orange and cyan lines show $1/J_R$ and $1/J_L$, respectively. The mean field is not exact because it does not capture the exclusion and the correlations imposed by it. Both the simulation and the mean field calculations show that the bulk (inverse output current) and the single particle (mean escape time) time scales are different and have different trends. The parameters used are: $r_{12}=r_{21}=1$, $r=1$, $r_L=1$, $r_R=0.1$, $m=1$ and $N=10$.}
\end{figure*}

For a channel of an arbitrary length, it is not possible to derive an analytical solution for the dynamics of a tagged particle. Therefore, the tagged particle dynamics is assumed to be affected only by the steady state population in the channel. This assumption is not exact, but for soft exclusion, where each site may include a few particles, it was shown to be a reasonable quantitative approximation. \cite{Zilman2009,Zilman2010}

To efficiently describe the dynamics of the system, it is convenient to employ a matrix representation. Using this approach, the corresponding equations can be written as:
\begin{equation}\label{eq:gcmr}
 \frac{d}{dt}|p(t)\rangle=\hat{U}^{ss}|p(t)\rangle,
\end{equation}
where  $|p(t)\rangle$ is the vector of stationary probabilities for different states, while $\hat{U}^{ss}$ describes the matrix consisting of transition rates. The details of the calculations and the matrix elements are fully explained in Appendix \ref{app:gc}.
The mean exit time  to the left is given in the matrix language as \begin{equation}\label{eq:gctl}
\overline{T}^{ss}_{\leftarrow}=\frac{r_L\left\langle 1\left\vert \left(
\left(\hat{U}^{ss}\right)^{-1}\right)^{2} \right\vert 2\right\rangle}{P_{\leftarrow}},
\end{equation}
where
\begin{equation}\label{eq:gcpl}
P_{\leftarrow}=-r_L \left\langle 1\left|\left(\hat{U}^{ss}\right)^{-1}\right|2\right\rangle.
\end{equation}
Similarly, the mean time to exit to the right is equal to
\begin{equation}\label{eq:gctr}
\overline{T}^{ss}_{\rightarrow}=\frac{r_R\left\langle N\left\vert \left(
\left(\hat{U}^{ss}\right)^{-1}\right)^{2} \right\vert 2\right\rangle}{P_{\rightarrow}},
\end{equation}
where
\begin{equation}\label{eq:gcpr}
P_{\rightarrow}=-r_L \left\langle N\left|\left(\hat{U}^{ss}\right)^{-1}\right|2\right\rangle.
\end{equation}
(see Appendix \ref{app:gc} for the detailed expressions and derivations).

In Figure \ref{fig:TimesGCrR0p1}, the mean exit times to the right and to the left, along with the corresponding inverse currents, are presented as a function of the incoming flux $f$. As $f$ increases, the exit current (in both directions) also increases. Naively, one would expect to see a corresponding decrease in the mean exit times. However, these times actually increase. 
Moreover, we again observe that the bulk time scales deduced from the exit currents are different from those deduced from the dynamics of the tagged particles.

One can notice that although our theoretical predictions agree with computer simulations, there are some deviations. These come from the fact that the mean-field approach does not capture the correlations in the steady state density and the dynamics of tagged particles. Due to the fact that our system is effectively one-dimensional, the correlations are expected to be strong. 

\section{\label{sec:sum}Summary and Conclusions}

In this paper, we developed a general theoretical framework to describe different time scales in complex dynamic systems with multiple outcomes. It is shown that for every exit, there are two time scales, the mean exit time and inverse exit current, that specify the dynamics of the system in this specific direction. Our theoretical arguments are explicitly illustrated by analyzing three different dynamic systems, including an enzyme with two substrates, a two-site channel with two exits, and a multi-site channel with two exits. Theoretical calculations for these systems were done using exact analytical calculations, a mean-field approximation, and kinetic Monte Carlo computer simulations. 

Our theoretical analysis shows that the two time scales may behave very differently, and this is the consequence of the existence of other exits in the system. This indicates that it is not correct to consider dynamics at each exit as independent from each other. In addition, it is argued that our theoretical calculations have a strong implication for the experimental studies of complex natural processes. This is because the mean exit times are typically determined from single-particle measurements, while the fluxes are obtained via bulk measurements. We conclude that both types of experimental measurements are needed in order to present a comprehensive description of the dynamics in such systems. It will be important to test experimentally our theoretical predictions. 

\begin{acknowledgments}
ABK acknowledges the support from the Welch Foundation (C-1559), from the NSF (CHE-1664218, CHE-1953453 and MCB-1941106), and from the Center for Theoretical Biological Physics sponsored by the NSF (PHY-1427654).
\end{acknowledgments}

\section*{Data Availability Statement}
The data that support the findings of this study are available from the corresponding author upon reasonable request.

\appendix

\section{Simple kinetic proofreading scheme\label{app:kpr}}
The molecular fluxes can be determined using the forward master equations. Since we consider the stationary dynamics, it can be assumed that as soon as the system reaches the state $R$ or $W$, it immediately resets to the state $E$ (see Fig. 1). One can the define $P_{E}$, $P_{ER}$ and $P_{EW}$ as stationary probabilities to find the system in the corresponding states $E$, $ER$ and $EW$. The stationary dynamics at state $E$ is described as the balance of the fluxes into this state and out of this state,
\begin{equation}\label{eq8}
    0=(u_{1}+w_{1})P_{ER}+(a_{1}+b_{1})P_{EW} - (u_{0}+w_{0}) P_{E},
\end{equation}
while for states $ER$ and $EW$, we have 
\begin{equation}\label{eq9}
0=u_{0}P_{E}- (u_{1}+w_{1}) P_{ER},
\end{equation}
and 
\begin{equation}\label{eq10}
0=a_{0}P_{E}- (a_{1}+b_{1}) P_{EW},
\end{equation}
respectively. In addition, the normalization requires that
\begin{equation}\label{eq11}
    P_{E}+P_{ER}+P_{EW}=1.
\end{equation}

Solving Eqs. (\ref{eq9}), (\ref{eq10}), and (\ref{eq11}) leads to explicit expressions for the stationary probabilities of different chemical states:
\begin{equation}
    P_{E}=\frac{(u_{1}+w_{1})(a_{1}+b_{1})}{(u_{1}+w_{1})(a_{1}+b_{1})+u_{0}(a_{1}+b_{1})+a_{0}(u_{1}+w_{1})};
\end{equation}
\begin{equation}
    P_{ER}=\frac{u_{0}(a_{1}+b_{1})}{(u_{1}+w_{1})(a_{1}+b_{1})+u_{0}(a_{1}+b_{1})+a_{0}(u_{1}+w_{1})};
\end{equation}
and
\begin{equation}
    P_{EW}=\frac{a_{0}(u_{1}+w_{1})}{(u_{1}+w_{1})(a_{1}+b_{1})+u_{0}(a_{1}+b_{1})+a_{0}(u_{1}+w_{1})}.
\end{equation}
This allows us to estimate the molecular flux to make the right products $R$,
\begin{equation}\label{App_eq15}
    J_{R}=u_{1}P_{ER}=\frac{u_{0}u_{1}(a_{1}+b_{1})}{(u_{1}+w_{1})(a_{1}+b_{1})+u_{0}(a_{1}+b_{1})+a_{0}(u_{1}+w_{1})},
\end{equation}
and to make the wrong products $W$
\begin{equation}\label{App_eq16}
    J_{W}=a_{1}P_{EW}=\frac{a_{0}a_{1}(u_{1}+w_{1})}{(u_{1}+w_{1})(a_{1}+b_{1})+u_{0}(a_{1}+b_{1})+a_{0}(u_{1}+w_{1})}.
\end{equation}

To evaluate the probabilities to make $R$ and $W$ products ($\Pi_{R}$ and $\Pi_{W}$), and the mean exit times in the right and wrong direction ($T_{R}$ and $T_{W}$), the first-passage method will be utilized.\cite{van1992stochastic,redner-book} For convenience, let us focus on the exit in the $R$ direction. One can define the functions $F_{j}(t)$ ($j=EW$, $E$, $ER$ or $R$) as the  probability density functions to reach product $R$ at time $t$ for the first time before reaching product $W$, if initially the system started in state $j$. The time evolution of these first-passage probability functions is governed by backward master equations,
\begin{align}\label{eq17}
    \frac{dF_{E}(t)}{dt}=&u_{0} F_{ER}(t)+a_{0} F_{EW}(t)-(u_{0}+a_{0}) F_{E}(t);\nonumber \\
    \frac{dF_{EW}(t)}{dt}=&b_{1} F_{E}(t)-(a_{1}+b_{1}) F_{EW}(t);\\
    \frac{dF_{ER}(t)}{dt}=&u_{1} F_{R}(t)+w_{1} F_{E} (t)-(u_{1}+w_{1}) F_{ER}(t).\nonumber
\end{align}
In addition, we have $F_{R}(t)=\delta(t)$, which means that if the system starts in state $R$, the process is immediately at the terminal state, $R$. 

Eq. (\ref{eq17}) can be conveniently analyzed using the Laplace transform, $\widetilde{F}_{j}(s)=\int_{0}^{\infty} e^{-st} F_{j}(t)dt$, which modifies the original backward master equations into
\begin{align}
    (s+u_{0}+a_{0}) \widetilde{F}_{E}(s)&=u_{0} \widetilde{F}_{ER}(s)+a_{0} \widetilde{F}_{EW}(s); \nonumber \\
    (s+a_{1}+b_{1}) \widetilde{F}_{EW}(s)&=b_{1} \widetilde{F}_{E}(s); \\
    (s+u_{1}+w_{1}) \widetilde{F}_{ER}(s)&=u_{1} +w_{1} \widetilde{F}_{E}(s).\nonumber
\end{align}
Solving these equations leads to
\begin{widetext}
\begin{equation}
     \widetilde{F}_{E}(s) =
    \frac{u_{0}u_{1}(s+a_{1}+b_{1})}{(s+u_{1}+w_{1})(s+u_{0}+a_{0})(s+a_{1}+b_{1})-a_{0}b_{1}(s+u_{1}+w_{1})-u_{0}w_{1}(s+a_{1}+b_{1})}. 
\end{equation}
\end{widetext}
Now, we can explicitly evaluate the splitting probability and the mean exit time,
\begin{equation}\label{App_eq24}
    \Pi_{R} \equiv  \widetilde{F}_{E}(s=0)=\frac{u_{0}u_{1}(a_{1}+b_{1})}{u_{0}u_{1}(a_{1}+b_{1})+a_{0}a_{1}(u_{1}+w_{1})};
\end{equation}
and
\begin{eqnarray}\label{App_eq25}
    T_{R} &\equiv& - \frac{\left(\frac{d\widetilde{F}_{E}(s)}{ds}\right)_{s \rightarrow 0}}{\Pi_{R}}=  \\
    &&\frac{(a_{1}+b_{1})(u_{0}+u_{1}+w_{1})+a_{0}b_{1}\frac{u_{1}+w_{1}}{a_{1}+b_{1}} +a_{0}a_{1}}{u_{0}u_{1}(a_{1}+b_{1})+a_{0}a_{1}(u_{1}+w_{1})}.\nonumber
\end{eqnarray}

A similar analysis can be done for the exit dynamics in the direction of the wrong products. Here we derive the following expressions,
\begin{equation}\label{App_eq26}
    \Pi_{W} =\frac{a_{0}a_{1}(u_{1}+w_{1})}{u_{0}u_{1}(a_{1}+b_{1})+a_{0}a_{1}(u_{1}+w_{1})};
\end{equation}
and
\begin{equation}\label{App_eq27}
    T_{W}=\frac{(u_{1}+w_{1})(a_{0}+a_{1}+b_{1})+u_{0}w_{1}\frac{a_{1}+b_{1}}{u_{1}+w_{1}} +u_{0}u_{1}}{u_{0}u_{1}(a_{1}+b_{1})+a_{0}a_{1}(u_{1}+w_{1})}. 
\end{equation}

\section{Two-site system\label{app:2s}}

\subsubsection{\label{sssec:2sprob}The probabilities and currents}
The dynamic of the probabilities of the two-site model considered in Sub Sec. \ref{subsec:2s} may be described by the following set of equations:%
\begin{align}
    \frac{dp_{00}}{dt}=&-fp_{00}+r_Lp_{10}+r_Rp_{01};\nonumber\\
    \frac{dp_{10}}{dt}=&fp_{00}-\left(r_L+r_{12}\right)p_{10}+r_{21}p_{01}+r_Rp_{11};\nonumber\\
    \frac{dp_{01}}{dt}=&r_{12}p_{10}-\left(r_R+r_{21}+f\right)p_{01}+r_Lp_{11};\nonumber\\
    \frac{dp_{11}}{dt}=&fp_{01}-\left(r_L+r_R\right)p_{11}.
\end{align}
The steady state solution and the resulting steady state output currents are provided in Eqs. \eqref{eq:2sss}\textendash\eqref{eq:2sJ}.

In addition, one may find the steady state actual input flux:
\begin{equation}
    J_{in}=f\left(\leftidx{^{ss}}{p}{_{01}}+\leftidx{^{ss}}{p}{_{00}}\right).
\end{equation}
In the limit of a small impinging flux, the right and left output fluxes take the form:
\begin{align}
    J_R&=f\frac{r_{12} r_R }{r_{12}r_R+(r_{21}+r_R) r_L}+O(f^2)\nonumber \\
    J_L&=f\frac{r_L (r_{21} + r_R)}{r_{12}r_R+(r_{21}+r_R) r_L}+O(f^2).
\end{align}
In the limit of a jammed system, $f\to\infty$, the output currents are:
\begin{align}\label{eq:2sasymJ}
    J_R&\sim r_{12} r_R/(r_{12} + r_R)\nonumber \\
    J_L&\sim r_L.
\end{align}
The currents also provide the probabilities of each particle to exit to the right or left,
\begin{widetext}
\begin{align}
    p_R&=J_R/(J_R+J_L)=\frac{r_{12} r_R (f+r_L+r_R)}{f r_R (r_{12}+r_L)+f r_{12} r_L+(r_L+r_R) (r_R (r_{12}+r_L)+r_{21} r_L)},\nonumber \\
    p_L&= J_L/(J_R+J_L)=\frac{r_L (f (r_{12}+r_R)+(r_{21}+r_R) (r_L+r_R))}{f r_R (r_{12}+r_L)+f r_{12} r_L+(r_L+r_R) (r_R (r_{12}+r_L)+r_{21} r_L)}.
\end{align}
\end{widetext}
In the limit of a small input current, $f\to 0$, the probabilities are:
\begin{align}
    \lim_{f\to 0}p_R&=\frac{r_{12} r_R}{r_{21} r_L + r_R r_{12} + r_L r_R},\nonumber \\
    \lim_{f\to 0}p_L&=\frac{r_{21} r_L+r_L r_R}{r_{21} r_L + r_R r_{12} + r_L r_R}.
\end{align}
In the opposite limit of a crowded system, $f\to\infty$, the probabilities are:
\begin{align}
    \lim_{f\to \infty}p_R&=\frac{r_{12} r_R}{r_{12}(r_L+r_R)+r_L r_R},\nonumber \\
    \lim_{f\to \infty}p_L&=\frac{r_{12}r_L+r_L r_R}{r_{12}(r_L+r_R)+r_L r_R}.
\end{align}
Note that in this latter limit, site 1 is always occupied and the transition from site 2 to site 1 never takes place; therefore, the expressions are independent of the rate $r_{21}$.

\subsubsection{\label{sssec:2sfpt}The first passage time}

In order to calculate the mean first passage time, we write the backward master equations for two possible initial conditions, $10$\textendash site 1 is occupied and state 2 is not and $11$\textendash both sites are occupied. For the first passage time, we have to consider a tagged particle, and if there is another one in the system, it is not tagged. Therefore, in what follows, we will use $\bullet$ to denote a tagged particle, $\circ$ to denote an untagged particle, and $\square$ to denote an empty site.
In writing the backward master equation, we number the states as: $1$\textendash$\bullet\circ$; $2$\textendash$\bullet\square$; $3$\textendash$\square\bullet$; $4$\textendash$\circ\bullet$; and $e$ for the case where the tagged particle exits to the right.
Mathematically, we write the probability density of escaping to the right at time $t$, given that the system is in state $j$ and the initial condition is $1$, as $\leftidx{^1}{F}{_{R,j}}$. The general form of the backward master equation is:
\begin{equation}
\frac{dF_{i}}{dt}=-F_{i}{\displaystyle\sum\limits_{j}}
r_{ij}+{\displaystyle\sum\limits_{j}}
F_{j}r_{ji}
\end{equation}
where $r_{ij}$ is the transition rate from state $j$ to state $i$
For the initial condition $1$ ($\bullet\circ$), the set of equations is:
\begin{align}
\frac{d}{dt}\leftidx{^1}{F}{_{R,1}}  & =-\leftidx{^1}{F}{_{R,1}}(r_{L}+r_R)+\leftidx{^1}{F}{_{R,2}}r_{R},\nonumber\\
\frac{d}{dt}\leftidx{^1}{F}{_{R,2}}  & =-\leftidx{^1}{F}{_{R,2}}(r_{L}+r_{12})+\leftidx{^1}{F}{_{R,3}}r_{12},\nonumber\\
\frac{d}{dt}\leftidx{^1}{F}{_{R,3}}  & =-\leftidx{^1}{F}{_{R,3}}(r_{R}+r_{21}+f)+%
\leftidx{^1}{F}{_{R,2}}r_{21}+\leftidx{^1}{F}{_{R,4}}f+\leftidx{^1}{F}{_{e}}r_{R},\nonumber\\
\frac{d}{dt}\leftidx{^1}{F}{_{R,4}}  & =-\leftidx{^1}{F}{_{R,4}}(r_{L}+r_R)+\leftidx{^1}{F}{_{R,3}}r_{L}+\leftidx{^1}{F}{_{e}}r_{R}.%
\end{align}
The backward master equations are linear; therefore, we apply the Laplace transform and write the set of equations as:
\begin{align}
\leftidx{^1}{\widetilde{F}}{_{R,1}}\left(  r_{L}+r_{R}+s\right)  & =\leftidx{^1}{\widetilde{F}}{%
_{R,2}}r_{R},\nonumber\\
\leftidx{^1}{\widetilde{F}}{_{R,2}}\left(  r_{L}+r_{12}+s\right)  & =\leftidx{^1}{\widetilde
{F}}{_{R,3}}r_{12},\nonumber\\
\leftidx{^1}{\widetilde{F}}{_{R,3}}\left(  r_{R}+r_{21}+f+s\right)  & =\leftidx{^1}{\widetilde
{F}}{_{R,2}}r_{21}+\leftidx{^1}{\widetilde{F}}{_{R,4}}f+r_{R},\nonumber\\
\leftidx{^1}{\widetilde{F}}{_{R,4}}\left(  r_{L}+r_{R}+s\right)  & =\leftidx{^1}{\widetilde{F}}{_{R,3}}r_{L}+r_{R}.
\end{align}
Following our definitions, the probability density at $t=0$ is zero for all states except for state $e$ for which the Laplace transform was considered explicitly ($\leftidx{^1}{F}{_{e}}(t)=\delta(t)$).%
The solution for the Laplace transform of the first passage time to the right, given that the initial condition is $1$, is:%
\begin{widetext}
\begin{equation}
\leftidx{^1}{\widetilde{F}}{_{R,1}}(s)=\frac{r_{12}r_{R}^{2}(f+r_{L}+r_{R}+s)}{(r_{L}%
+r_{R}+s)\left(  r_{21}\left(  r_{L}+s\right)  (r_{L}+r_{R}+s)+(r_{L}%
+r_{12}+s)(r_{R}+s)(f+r_{L}+r_{R}+s)\right)  }%
\end{equation}
\end{widetext}
The corresponding right exit probability (given the initial state, 1) is provided in Eq. \eqref{eq:2spir1}.
The mean escape time according to $F_{R,1}$ is given in Eq. \eqref{eq:2sTR1}.%

For the initial state $2$ ($\bullet\square$), the backward master equation is:
\begin{align}
\frac{d}{dt}\leftidx{^2}{F}{_{R,2}}  & =-\leftidx{^2}{F}{_{R,2}}(r_{L}+r_{12})+\leftidx{^2}{F}{_{R,3}}r_{12},\nonumber\\
\frac{d}{dt}\leftidx{^2}{F}{_{R,3}}  & =-\leftidx{^2}{F}{_{R,3}}(r_{R}+r_{21}+f)+\leftidx{^2}{F}{_{R,2}}r_{21}+\leftidx{^2}{F}{_{R,4}}f+\leftidx{^2}{F}{_{e}}r_{R},\nonumber\\
\frac{d}{dt}\leftidx{^2}{F}{_{R,4}}  & =-\leftidx{^2}{F}{_{R,4}}(r_{L}+r_R)+\leftidx{^2}{F}{_{R,3}}r_{L}+\leftidx{^2}{F}{_{e}}r_{R}.%
\end{align}
Using the same approach as before, the Laplace transform of the equations reads:%
\begin{align}
\leftidx{^2}{\widetilde{F}}{_{R,2}}\left(  r_{L}+r_{12}+s\right)  & =\leftidx{^2}{\widetilde{F}}{_{R,3}}r_{12},\nonumber\\
\leftidx{^2}{\widetilde{F}}{_{R,3}}\left(  r_{R}+r_{21}+f+s\right)  & =\leftidx{^2}{\widetilde{F}}{_{R,2}}r_{21}+ \leftidx{^2}{\widetilde{F}}{_{R,4}}f+r_{R},\nonumber\\
\leftidx{^2}{\widetilde{F}}{_{R,4}}\left(  r_{L}+r_{R}+s\right)  & =\leftidx{^2}{\widetilde{F}}{_{R,3}}r_{L}+r_{R}.%
\end{align}
The solution of interest is%
\begin{widetext}
\begin{equation}
\leftidx{^2}{\widetilde{F}}{_{R,2}}=-\frac{r_{12}r_{R}(f+r_{L}+r_{R}+s)}{r_{12}r_{21}%
(r_{L}+r_{R}+s)+(r_{12}+r_{L}+s)(fr_{L}-(f+r_{21}+r_{R}+s)(r_{L}+r_{R}+s).)}.%
\end{equation}
\end{widetext}
The corresponding right exit probability (given initial state 2) is given in Eq. \eqref{eq:2spir2}.%
The corresponding mean time is given in Eq. \eqref{eq:2sTR2}.

The results above can be combined to determine the mean right escape time (considering the proper average over the two possible initial conditions), as appears in Eq. \eqref{eq:2sTR}.

\section{General channel\label{app:gc}}

\subsubsection{\label{sssec:popdyn}Population dynamics and steady state}
The equations describing the dynamics of the population are:
\begin{widetext}
\begin{align}\label{eq:pdyn}
    \frac{dn_k}{dt}&=-rn_k\left(1-\frac{n_{k+1}}{m}\right)-rn_k\left(1-\frac{n_{k-1}}{m}\right)+r\left(n_{k+1}+n_{k-1}\right)\left(1-\frac{n_k}{m}\right) \ \ \ \ \ 3\leq k \leq N-2, \nonumber \\
    \frac{dn_1}{dt}&=-\left(r_L+r_{12}\left(1-\frac{n_2}{m}\right)\right)n_1+r_{21}n_2, \nonumber \\
    \frac{dn_2}{dt}&=\left(f+r_{12}n_1+rn_3\right)\left(1-\frac{n_2}{m}\right)-\left(r_{21}+r\left(1-\frac{n_3}{m}\right)\right)n_2, \nonumber \\
    \frac{dn_{N-1}}{dt}&=\left(r_{12}n_N+rn_{N-2}\right)\left(1-\frac{n_{N-1}}{m}\right)-\left(r_{21}+r\left(1-\frac{n_{N-3}}{m}\right)\right)n_{N-2}, \nonumber \\
    \frac{dn_{N}}{dt}&=-\left(r_R+r_{12}\left(1-\frac{n_{N-1}}{m}\right)\right)n_N+r_{21}n_{N-1}.
\end{align}
\end{widetext}
The mean field steady state solution is provided in Eqs. \eqref{eq:sspan}\textendash\eqref{eq:sspan3}.

\subsubsection{Single particle using mean field in the jammed regime}

In the mean field approximation, the dynamics of the single particle is assumed to be affected by the steady state population only through the modification of the transition rates. The rates are assumed to be affected only by the mean density of states; therefore, the correlations due to the exclusion are neglected.
The probabilities of the tagged particle are described by the following equations \cite{redner-book,gardiner-book,bezrukov-sites-2005,Zilman2009,Zilman2010}:
\begin{widetext}
\begin{eqnarray}\label{eq:kinetics_spj}
\frac{d}{dt}{p}_k(t)=r\left(1-\frac{n_k^{ss}}{m}\right)(p_{k-1}+p_{k+1})-rp_k\left(2-\frac{n_{k+1}^{ss}}{m}-\frac{n_{k-1}^{ss}}{m}\right)\;\;\text{for}\;\;2<k<N-1
\end{eqnarray}
Taking into account the solution for the steady state population, \eqref{eq:sspan}, we can rewrite the equation as:
\begin{eqnarray}\label{eq:kinetics_spj2}
\frac{d}{dt}p_k(t)=
r\left(1-\frac{n_k^{ss}}{m}\right)\left(p_{k-1}+p_{k+1}-2p_k\right)=
r\left(A-kB\right)\left(p_{k-1}+p_{k+1}-2p_k\right)\;\;\text{for}\;\;2<k<N-1
\end{eqnarray}
The boundary conditions are written as
\begin{align}\label{eq:kinetics_sp_bcj}
\frac{d}{dt}{p}_1(t)&=-\left(r_L+r_{12}\left(1-\frac{n^{ss}_2}{m}\right)\right)p_1+r_{21}p_2=-\left(r_L+r_{12}\left(A-2B\right)\right)p_1+r_{21}p_2; \\
\frac{d}{dt}{p}_2(t)&=-\left(r\left(1-\frac{n^{ss}_3}{m}\right)+r_{21}\right)p_2+r_{12}\left(1-\frac{n^{ss}_2}{m}\right)p_1+r\left(1-\frac{n^{ss}_2}{m}\right)p_{3}\nonumber \\
&=-\left(r\left(A-3B\right)+r_{21}\right)p_2+r_{12}\left(A-2B\right)p_1+r\left(A-2B\right)p_{3};\nonumber \\
\frac{d}{dt}{p}_{N-1}(t)&=-\left(r_{21}+r\left(1-\frac{n^{ss}_{N-2}}{m}\right)\right)p_{N-1}+r\left(1-\frac{n^{ss}_{N-1}}{m}\right)p_{N-2}+r_{12}\left(1-\frac{n^{ss}_{N-1}}{m}\right)p_{N}\nonumber \\
&=-\left(r_{21}+r\left(A-(N-2)B\right)\right)p_{N-1}+r\left(A-(N-1)B\right)p_{N-2}+r_{12}\left(A-(N-1)B\right)p_{N};\nonumber \\
\frac{d}{dt}{p}_N(t)&=-(r_R+r_{12}\left(1-\frac{n^{ss}_{N-1}}{m}\right))p_N+r_{21}p_{N-1}=-(r_R+r_{12}\left(A-(N-1)B\right))p_N+r_{21}p_{N-1}.\nonumber
\end{align}
\end{widetext}
Using a matrix representation, the dynamics may be written as:
\begin{equation}
 \frac{d}{dt}|p(t)\rangle=\hat{U}^{ss}|p(t)\rangle.
\end{equation}
The matrix elements for $2<k<N-1$ are given by:
\begin{align}
U^{ss}_{k,k}=-2r(1-n^{ss}_k/m)=-2r(A-kB);\nonumber \\
U^{ss}_{k,k\pm 1}=r(1-n^{ss}_k/m)=r(A-kB).
\end{align}
The boundary conditions are represented by the following matrix elements:
\begin{align}
&U^{ss}_{1,1}=-r_{12}\left(1-\frac{n^{ss}_2}{m}\right)-r_L=-r_{12}(A-2B)-r_L;\nonumber \\
&U^{ss}_{2,2}=-r\left(1-\frac{n^{ss}_3}{m}\right)-r_{21}=-r(A-3B)-r_{21};\nonumber \\
&U^{ss}_{N-1,N-1}=-r\left(1-\frac{n^{ss}_{N-2}}{m}\right)-r_{21}=-r(A-(N-2)B)-r_{21};\nonumber \\
&U^{ss}_{N,N}=-r_{12}\left(1-\frac{n^{ss}_{N-1}}{m}\right)-r_R=-r_{12}\left(A-(N-1)B\right)-r_R; \nonumber \\
&U^{ss}_{1,2}=r_{21};\nonumber \\
&U^{ss}_{2,1}=r_{12}\left(1-\frac{n^{ss}_{2}}{m}\right)=r_{12}(A-2B);\nonumber \\
&U^{ss}_{2,3}=r\left(1-\frac{n^{ss}_2}{m}\right)=r(A-2B);\nonumber \\
&U^{ss}_{N-1,N-2}=r\left(1-\frac{n^{ss}_{N-1}}{m}\right)=r(A-(N-1)B);\nonumber \\
&U^{ss}_{N-1,N}=r_{12}\left(1-\frac{n^{ss}_{N-1}}{m}\right)=r_{12}(A-(N-1)B);\nonumber \\
&U^{ss}_{N,N-1}=r_{21}.
\end{align}
The average exit times to the right/left and the corresponding probabilities are provided in Eqs. \eqref{eq:gctl}\textendash\eqref{eq:gcpr}.

In order to obtain an explicit expression for $\overline{T}^{ss}_{\leftarrow}$, we define
\begin{equation}
|\alpha\rangle=D  \left(  U^{ss}\right)  ^{-1}\left\vert
2\right\rangle;
\end{equation}
\begin{equation}
\langle \leftidx{_L}{Q}|=D  \left\langle 1\right\vert \left(  U^{ss}\right)
^{-1};
\end{equation}
and
\begin{equation}
\langle \leftidx{_R}{Q}|=D  \left\langle N\right\vert \left(  U^{ss}\right)
^{-1};
\end{equation}
In the equations above, we introduced the notation for the determinant of $U^{ss}$, $D\equiv det\left(U^{ss}\right)$.
The elements of these vectors are given by
\begin{widetext}
\begin{align}
&\alpha_1=\alpha_0\left(rr_{12}\left(A^2+2(N-1)B^2-AB(N+1)\right)+rr_R\left(A-2B\right)+(N-3)r_Rr_{21}\right);\nonumber \\
&\alpha_2=\alpha_1\left(Ar_{12}-2Br_{12}+r_L\right)/r_{21};\nonumber \\
&\alpha_k=\alpha_0\left(\left(A^2+k(N-1)B^2-AB(N+k-1)\right)rr_{12}+rr_R\left(A-kB\right)+(N-k-1)r_Rr_{21}\right);\ \ 2<k<N-1 \nonumber \\
&\alpha_{N-1}=\alpha_N\left(Ar_{12}-(N-1)Br_{12}+r_R\right)/r_{21};\nonumber \\
&\alpha_N=\alpha_0rr_{21}\left(  A-B\right)\left(  A-2B\right)\left(Ar_{12}-2Br_{12}+r_L\right),
\end{align}
where $\alpha_0=r^{N-4}r_{21}\left(-1\right)^{N-1}\left({\displaystyle\prod\limits_{k=3}^{N-2}}
\left(  A-kB\right)\right)$. The elements of the vector $\leftidx{_L}{Q}$ are given by:
\begin{align}
&\leftidx{_L}{Q}{_1}=\alpha_0\left(rr_{12}\left(A^2+2(N-1)B^2-AB(N+1)\right)+rr_R\left(2A-(N+1)B\right)+(N-3)r_Rr_{21}\right);\nonumber \\
&\leftidx{_L}{Q}{_2}=\alpha_0\left(rr_{12}\left(A^2+2(N-1)B^2-AB(N+1)\right)+rr_R\left(A-2B\right)+(N-3)r_Rr_{21}\right);\nonumber \\
&\leftidx{_L}{Q}{_{k}}=\alpha_0\frac{A-2B}{A-kB}\left(rr_{12}\left(A^2+k(N-1)B^2-AB(N+k-1)\right)+rr_R\left(A-kB\right)+(N-k-1)r_Rr_{21}\right);\ \ 2<k<N-1\nonumber \\
&\leftidx{_L}{Q}{_{N-1}}=\leftidx{_L}{Q}{_{N}}\left(1+\frac{r_R}{\left(A-(N-1)B\right)r_{12}}\right)\nonumber \\
&\leftidx{_L}{Q}{_{N}}=\alpha_0rr_{12}\left(A-2B\right)\left(A-(N-1)B\right).
\end{align}
The elements of the vector $\leftidx{_R}{Q}$ are given by:
\begin{align}
&\leftidx{_R}{Q}{_1}=\leftidx{_L}{Q}{_{N}}=\alpha_0rr_{12}(A-2B)(A-(N-1)B);
\nonumber \\
&\leftidx{_R}{Q}{_2}=\alpha_0r\left(A-\left(N-1\right)B\right)\left(\left(A-2B\right)r_{12}+r_L\right);\nonumber \\
%
&\leftidx{_R}{Q}{_{k}}=\frac{\alpha_0\left(A-(N-1)B\right)}{A-kB}\left(rr_{12}\left(A^2+2kB^2-AB(k+2)\right)+rr_L\left(A-kB\right)+(k-2)r_Lr_{21}\right);\ \ 2<k<N-1\nonumber \\
&\leftidx{_R}{Q}{_{N-1}}=\alpha_0\left(rr_{12}\left(A^2+2(N-1)B^2-AB(N+1)\right)+rr_L\left(A-(N-1)B\right)+(N-3)r_Lr_{21}\right);\nonumber \\
&\leftidx{_R}{Q}{_{N}}=\alpha_0\left(rr_{12}\left(A^2+2(N-1)B^2-AB(N+1)\right)+rr_L\left(2A-(N+1)B\right)+(N-3)r_Lr_{21}\right).
\end{align}
\end{widetext}
In the above expressions, $A$ and $B$ are set by the solution of the set of equations \eqref{eq:sspan3}.
The mean escape time to the left is then
\begin{equation}
\overline{T}^{ss}_{\leftarrow}P_{\leftarrow}=\frac{r_{L}}{D^{2}}%
{\displaystyle\sum\limits_{k=1}^{N}}
\leftidx{_L}{Q}{_{k}}\alpha_{k},
\end{equation}
and similarly,
\begin{equation}
\overline{T}^{ss}_{\rightarrow}P_{\leftarrow}=\frac{r_{R}}{D^{2}}%
{\displaystyle\sum\limits_{k=1}^{N}}
\leftidx{_R}{Q}{_{k}}\alpha_{k}.
\end{equation}
Using the notations above we can express $D$ as:
\begin{widetext}
\begin{equation}
D=\alpha_0(A-B)(A-2B)\left(rr_{12}(r_L+r_R)(A-2B)(A-(N-1)B)+r_Lr_R\left(r(2A-(N+1)B)+(N-3)r_{21}\right)\right).
\end{equation}
\end{widetext}
The detailed expressions are cumbersome and, therefore, are not provided in detail here.\\
\nocite{*}

\bibliography{aipsamp}

\end{document}